\documentclass{iaus}
\usepackage{graphicx}

\title[Temperature Variations in PNe]
{Temperature Variations and Chemical Abundances in Planetary Nebulae}

\author[Peimbert \& Peimbert]
{Manuel Peimbert$^1$ \and Antonio Peimbert$^1$}
 
\affiliation{$^1$ Instituto de Astronom\'{\i}a,
  Universidad Nacional Aut\'onoma de M\'exico, \break
  Apdo. postal 70-264, M\'exico D.F. 04510, M\'exico \break
  email: peimbert@astroscu.unam.mx}

\pubyear{2006}
\volume{234}
\pagerange{1-8}
\date{May 3, 2006}
\jname{Planetary Nebulae in our Galaxy and Beyond}
\editors{M. J. Barlow \& R. H. M\'endez, eds.}
\begin{document}

\maketitle

\begin{abstract}

In this review we present a brief discussion on the observational
evidence in favor of the presence of temperature variations, and
conclude that many planetary nebulae show spatial temperature
variations that are larger than those predicted by 1D static
chemically homogeneous photoionization models.  To determine accurate
chemical abundances it is necessary to know the cause of these
temperature variations and several possibilities are discussed. The
importance of this problem is paramount to test the models of stellar
evolution of low and intermediate mass stars and of the chemical
evolution of galaxies. We conclude that the proper abundances for
chemically homogeneous PNe are those derived from recombination lines,
while for the two-abundance nebular model the proper heavy element
abundances relative to hydrogen are those derived from visual and UV
collisionally excited lines adopting the $t^2$ values derived from
$T_e$([O~III]) and $T_e$(Balmer).

\end{abstract}

\firstsection 
\section{Overview}
Recent reviews on the presence of temperature variations in PNe have been presented
by Esteban (2002), Liu (2003, 2006), and Torres-Peimbert \& Peimbert (2003). Of 
the well observed PNe about one third can be fitted reasonably well by 1D static chemically 
homogeneous photoionization models, but two thirds show temperature variations that 
are substantially larger than those predicted by simple photoionization models. In 
this paper we review further evidence in favor of the presence of large temperature 
variations in PNe. We discuss possible causes for these variations and their effect 
on the determination of chemical abundances of PNe. We discuss the proper procedure 
to obtain accurate abundances for chemically homogeneous and for chemically 
inhomogeneous PNe.

\section{Temperature determinations}

Peimbert (1967, 1971) found that the determinations of $T_e$([O~III]), the temperatures 
based on the $I(4363)/I(5007)$ ratio, are considerably larger than the determinations 
of $T_e$(Balmer), the temperatures based on the intensity ratio of the Balmer continuum to a 
Balmer recombination line; he interpreted this result as being due to the presence of
temperature variations over the observed volume. To study this problem, Peimbert defined the 
mean square temperature variation, $t^2$; typical simple photoionization models yield $t^2$ 
values in the 0.003 - 0.015 range. Peimbert and collaborators, also developed equations 
to derive the abundances for chemically homogeneous nebulae with $t^2 > 0.000$ (Peimbert 1967; 
Peimbert \& Costero 1969, Ruiz et al. 2003, Peimbert et al. 2004). The large 
differences between $T_e$([O~III]) and $T_e$(Balmer) have been confirmed by several authors 
for a large number of PNe (e.g. Liu \& Danziger 1993; Zhang, Liu, \& Wesson 2004).

Peimbert, Storey, \& Torres-Peimbert (1993) based on the computations
by Storey (1994) were the first to obtain larger O$^{++}$/H$^+$ values
from oxygen recombination lines than from collisionally excited lines
under the assumptions of $t^2$ = 0.000 and of chemical homogeneity.
Most PNe show this difference which has been usually called the
abundance discrepancy factor defined by: adf(O$^{++}$/H$^+$) =
(O$^{++}$/H$^+$)$_{RL}$ / (O$^{++}$/H$^+$)$_{CEL}$, (e. g. Liu 2006,
and references therein). The adf(O$^{++}$/H$^+$) value is larger than
predicted by simple photoionization models for about two thirds of the
well observed PNe.

A similar abundance difference for C$^{++}$/H$^+$ has been obtained by
Peimbert, Torres-Peimbert, \& Luridiana (1995a) based manly on the
line intensities compilation by Rola \& Stasi\'nska (1994). Peimbert
et al. (1995a) compared the C$^{++}$/H$^+$ abundances derived from the
C~II $\lambda$ 4267 recombination line with those derived from the
C$^{++}$ $\lambda\lambda$ 1906, 1909 collisionally excited lines.
Again about one third of the adf(C$^{++}$/H$^+$) of the well observed
PNe might be explained by simple photoionization models but two thirds
present values too large to be reproduced by these models.

Zhang et al. (2005) have obtained large $T_e$([O~III]) $-$ $T_e$(He~I) 
differences for 48 PNe that con not be explained by simple photoionization models.
In section 4.2 we discuss their results.

\section{Possible sources of temperature variations}

Torres-Peimbert \& Peimbert (2003) presented seven mechanisms as possible sources of temperature
variations, in what follows we will mention additional results in favor of some of these
mechanisms. 

{\it Deposition of mechanical energy:} The central stars of PNe inject
mechanical energy into the expanding shells by means of stellar winds,
bipolar flows, multipolar flows, and asymmetrical ejections, these
processes produce shocks, turbulence, and an increase of the expansion
velocity of the shell with time.  These processes are more important
for some objects than for others and might be responsible for the
spread in the observed $t^2$ values. Guerrero et al. (2005) and
Guerrero, Chu, \& Gruendl (2006) found that the following PNe are
strong X-ray emitters: BD+30$^o$~3639, NGC~40, NGC~2392, NGC~3242,
NGC~6543, NGC~7009, and NGC~7027, they also argue that this emission
is due to shocks produced by fast winds or bipolar flows. Rowlands,
Houck, \& Herter (1994) derived $T_e$([Ne V]) $\sim$ 50,000 K for
NGC~6302 and NGC~6537, they also computed photoionization models and
were not able to produce temperatures higher than 20,000~K and reached
the conclusion that these temperatures were indicative of shock
heating. Peimbert et al. (1995a) found that the following PNe with
large velocity dispersions also show large temperature variations:
NGC~2392, NGC~2371-2, NGC~2818, NGC~6302, and Hu~1-2, the last four
are bipolar Type~I PNe. Medina, Pe\~na, \& Stasi\'nska (2006) found,
from a sample of 47 PNe, that the velocity of expansion of the shell
increases with age indicators, for example $\left<v_{exp}\right>$ is
larger for low density nebulae and it is also larger for nebulae
with higher temperature stars. Furthermore by studing the line
profiles Medina et al. (2006) find that a substancial fraction of the
material shows turbulent motions.

{\it Chemical inhomogeneities:} See section 4.2 and the review by Liu (2006).

{\it Time dependent ionization:} When a photoionization front passes through a nebula it heats the gas
above the steady state value and some time is needed to reach thermal equilibrium.
In the presence of localized density variations or a density gradient this process 
produces large temperature variations and might explain the presence of hot external
halos in PNe (e.g. Tylenda 2003, Sandin et al. 2006). When the stellar ionizing flux 
decreases, or the density distribution along the line of sight changes, the outer regions 
of a nebula might become isolated from the stellar radiation field
and will cool down before fully recombining, creating cold partially ionized outer
regions, this might explain the low $T_e$(Balmer) values derived by Luo \& Liu (2003) in 
the outer regions of NGC~7009.

{\it Density variations:} Extreme density variations are present in most PNe, as can be seen
from optical images. For steady state photoionization models density variations are not very 
important, but for time dependent processes the regions of higher density will reach equilibrium
sooner than those of lower density.

{\it Deposition of magnetic energy:} No specific models have been proposed yet for this mechanism.

{\it Dust heating:} Stasi\'nska \& Szczerba (2001) have analyzed the
effects of photoelectric heating by dust grains in photoionization
models of PNe. This effect might be important in nebulae with large
density variations. This suggestion has not been tested yet for a 
specific model on any given PN.

{\it Shadowed regions:} Due to the presence of molecular globules inside NGC 7293 and NGC 6720,
two nearby PNe, their presence is expected in many PNe. According to Huggins \& Frank (2006),
the covering factor of the globules in NGC 7293 amounts to about 5\%. The ionization of
the gas shadowed by the globules will be produced by diffuse radiation, and consequently,
the temperature of the shadowed gas will be a few thousand degrees lower than that of
the material that is directly ionized by the central star (Mathis 1976). This mechanism
alone might produce $t^2$ values around 0.01 in PNe of the type of NGC 7293 and NGC 6720.

To discriminate among the different possibilities, it is important to
understand the signature of each process on the temperature and
density distributions. Mechanical energy deposition, an increase in the local
ionizing flux, and magnetic energy deposition will produce localized
high-temperature regions relative to simple photoionization predictions;
while shadowed regions, the decrease of the local ionizing flux, and the
presence of metal-rich inclusions will produce localized low-temperatures
regions relative to simple photoionization predictions.

\section{Forbidden or recombination line abundances?}

\subsection{Chemically homogeneous case}

Abundances correspond to those derived from recombination lines, if forbidden
lines are used a $t^2$ different from 0.000 has to be adopted. In what follows
we present evidence in favor of chemical homogeneity for most PNe.

From chemical evolution models of the Galaxy it has been found that in the solar
vicinity about half of the C enrichment of the ISM is due to low and intermediate
mass stars that end their lives as white dwarfs, and the rest is due to SN of
Type~II (e. g. Carigi et al. 2005). Moreover according to other models most of
the C enrichment is due to low and intermediate mass stars (e. g. Matteucci 2006).
Carigi (2003) has shown that the C/H values derived from C~II recombination lines
are in agreement with these models, while the C/H  values derived from the 1906 and
1909 collisionally excited lines, under the assumption of $T_e$([O~III]) and $t^2 = 0.000$,
imply lower C yields than those needed by the Galactic chemical evolution models.

\begin{table}\def~{\hphantom{0}}
  \begin{center}
  \caption{Stellar and Nebular Abundances for NGC 6543}
  \label{tab:6543}
  \begin{tabular}{lcccc}\hline
$N$($X$)/$N$(H) & CELs          &   RLs           & Central Star  & Solar Values  \\
    \hline
He/H            & ...           & $0.117\pm0.004$ & $0.11\pm0.01$ & ...           \\
12 + log C/H    & 8.2-8.5       & $8.90\pm0.10$   & $9.0\pm0.1$   & $8.39\pm0.05$ \\
12 + log O/H    & $8.86\pm0.10$ & $9.15\pm0.10$   & $9.1\pm0.1$   & $8.66\pm0.05$ \\
  \hline 
  \end{tabular}
  
  {\it References:} C/H Rola \& Stasi\'nska (1994), Peimbert, et al.
  (1995a), Wesson \& Liu (2004); O/H(FL) and all RLs Wesson \& Liu
  (2004), central star Georgiev et al.(2006); solar values Asplund,
  Grevesse, \& Sauval (2005).

  \end{center}
\end{table}

Esteban et al.(2005), based on recombination lines of C~II and O~II of Galactic
H~II regions, have determined for the ISM of the solar vicinity that 12 + log O/H =
$8.77\pm0.05$ and 12 + log C/H = $8.67\pm0.07$. These values are in excellent agreement
with the Asplund, Grevesse, \& Sauval (2005) solar values (see Table 1), considering that since
the Sun was formed the increase in the ISM abundances of these elements has been
of 0.13 dex in O/H and 0.29 dex in C/H; the increases in C/H and O/H are those
predicted by Galactic chemical evolution models by Carigi et el. (2005).

In Table 1 we present the stellar abundances of NGC~6543 based on a non-LTE
model by Goergiev et al. (2006), and compare them with those derived from
recombination and forbidden lines of the gaseous nebula. The recombination
O$^{++}$ abundance is based only on the multiplet 1 of O~II. The agreement
between the nebular recombination line abundances and the stellar abundances is
excellent, while the forbidden line abundances are about a factor of two to four
smaller than the stellar ones. This result is in favor of the idea that the
recombination line abundances are the proper ones for this object.

The O/H value for the central star of NGC~6543 is higher than the
solar value, see Table 1. There are three factors that might help to
explain the difference: a) the Sun was formed 4.5 Gyr ago and, as
mentioned above, the C/H and O/H values of the ISM have increased
during this period, therefore we would expect PNe with progenitor
masses greater than 2M$_{\odot}$ to have been formed when the ISM had
abundances greater than solar; b) a fraction of the H has been
converted into He increasing the C/H and O/H ratios, this is a small
effect and is in the 0.02 to 0.06~dex range; and c) some intermediate
mass star models predict an increase in the O/H ratio, Marigo,
Bressan, \& Chiosi (1996) obtain an increase of about 0.2~dex in the
O/H ratio for stellar models in the 1.83 to 2.5M$_{\odot}$ mass range
with Z=0.008.  It is clear that accurate abundances for the
atmospheres of the central stars of PNe are needed to test the models
of stellar evolution.

Liu et al. (2001), presented a strong correlation between
the adf(O$^{++}$/H$^+$) and $T_e$([O~III]) $-$ $T_e$(Balmer) and mention
that this correlation strongly supports the idea that temperature variations are real.
Similarly, from the adf(C$^{++}$/H$^+$) values by Peimbert et al.(1995a) and others
in the literature and the $T_e$(Balmer) values by Zhang et al.(2004), a strong correlation
between the adf(C$^{++}$) and $T_e$([O~III]) $-$ $T_e$(Balmer) is found, result that also supports
the presence of temperature variations.

From the relative intensities of the lines of multiplet 1 of O~II it is possible to 
obtain $N_e$(O~II) (Ruiz et al. 2003; Peimbert \& Peimbert 2005, Bastin \& Storey 2006).
The $N_e$(O~II) values can be compared with the $N_e$(Balmer) values obtained from 
Zhang et al. (2004). In Table~2 we present the densities obtained from [Cl~III], O~II, 
and H~I lines for five PNe with relatively high adf(O$^{++}$) values (see Liu 2006, 
and references therein). The $N_e$(O~II) values were obtained from the equations 
presented by Peimbert \& Peimbert (2005), these equations were derived from a calibration based 
on H~II regions. The $N_e$(O~II) values presented in Table~2 probably are slightly higher than the real
ones because the temperature of the H~II regions used for the calibration are larger than those 
of the PNe in Table~2, therefore the $N_e$(O~II)values presented in Table 2 
should probably be reduced by 0.1 to 0.2 dex due to the temperature difference.
The equations determined by Ruiz et al., that are based on a fit to PNe and H~II regions, 
yield $N_e$(O~II) values about 0.25 dex smaller than those presented in Table~2.
The $N_e$(O~II) values are in good agreement with the $N_e$(Balmer) values 
supporting the idea that these objects are chemically homogeneous. The $N_e$([Cl~III]) 
values are smaller than the $N_e$(Balmer) values, which is 
expected in the presence of a medium with density and temperature variations. 

\begin{table}\def~{\hphantom{0}}
  \begin{center}
  \caption{Electron Densities (cm$^{-3}$)}
  \label{tab:ne}
  \begin{tabular}{lccc}\hline
Object      & log $N_e$[Cl~III] & log $N_e$(Balmer) &  log $N_e$(O~II) \\
    \hline
NGC~6153    & $3.6\pm0.2$       & $3.8\pm0.2$       & $3.9\pm0.3$      \\
NGC~6543    & $3.7\pm0.1$       & $3.8\pm0.2$       & $4.0\pm0.4$      \\
NGC~7009    & $3.5\pm0.2$       & $3.8\pm0.1$       & $4.2\pm0.3$      \\
M1-42       & $3.2\pm0.1$       & $3.7\pm0.2$       & $3.9\pm0.3$      \\
M2-36       & $3.7\pm0.1$       & $3.8\pm0.1$       & $4.0\pm0.3$      \\
  \hline
  \end{tabular}
  
$N_e$ [Cl~III], $N_e$(O~II) Peimbert \& Peimbert (2005, and references therein);
$N_e$ (Balmer) Zhang et al. (2004).

  \end{center}
\end{table}

Chemically inhomogeneous nebulae can be produced by H-poor stars that
eject material into H-rich nebulae. That is the case of A30 and A78
(Jacoby 1979; Hazzard et al. 1980; Jacoby \& Ford 1983; Manchado,
Potash \& Mampaso 1988; Wesson \& Liu 2003). This type of situation
might occur in those cases where the central star is H-poor.
According to Gorny \& Tylenda (2000) about 10\% of the central stars
of PNe are H-poor; while, from the results of the Sloan project, based
on 2065 DA and DB white dwarfs Kleinman et al. (2004) find that 1888
are non magnetic DAs and 177 are non magnetic DBs. From these numbers,
we conclude that about 10\% of Galactic PNe have a H-poor central
star, and might show He, C, and O rich inclusions in their expanding
shells.  We consider unlikely for PNe with H-rich central stars to
have significant amounts H-poor material in their associated nebulae.

\subsection{Chemically inhomogeneous case}

It has been proposed that many PNe are chemically inhomogeneous (e.g. Liu
2006 and references therein). In this proposal, the two-abundance nebular model, 
PNe present two components: a) the low density component, that has most of the 
mass and is relatively hot, emits practically all the intensity of the H lines and of
the forbidden lines in the visual and the UV, and part of the intensity of the He~I lines,
and b) the high density component, that has only a small fraction of the total mass, 
is relatively cool, H-poor, and rich in heavy elements, and emits part of the He~I and of
the recombination line intensities of the heavy elements but practically no H and no
heavy element collisionally excited lines.

In favor of the two-abundance nebular model is that it provides an explanation for the
observed $T_e$(Balmer) $-$ $T_e$(He~I) differences, but it does not
provide an explanation for the temperature variations responsible for
the difference between $T_e$([O~III]) and $T_e$(Balmer). The main evidence for the two abundance nebular 
model has been provided by
Zhang et al. (2005) who found an average difference of $T_e$(Balmer) $-$ $T_e$(He~I) = 4000~K
from the ratio of the $\lambda$ 6678 to $\lambda$ 7281 recombination lines of He~I in 48 PNe. 

The abundances of the low density component are the ones needed for
studies of the chemical evolution of galaxies and of low and
intermediate mass stars, therefore we will discuss its abundances. For
the low density component the forbidden lines with $t^2 = 0.000$
provide a lower limit and the recombination lines provide an upper
limit to the real abundances relative to H. The $t^2$ formalism
applies to any type of gaseous nebulae, but the equations to derive
abundance ratios by (Peimbert \& Costero 1969, Ruiz et al. 2003,
Peimbert et al. 2004) assume chemical homogeneity; in the
two-abundance model, the low density component is chemically
homogeneous, thus its abundance can be computed from this formalism
using a $t^2$ that is representative of this volume; since we don't
expect the high density component to have neither relevant hydrogen
emission (it contains very little hydrogen) nor relevant [O~III]
$\lambda\lambda$ 4363, 5007 emission (it is too cold), the $t^2$ that
can be determined from $T_e$(Balmer) and $T_e$([O~III]) will only be
representative of this volume and the abundances determined from
H$\beta$, [O~III] $\lambda$ 5007, and this $t^2$ will have no
contamination from the emission of the very small high density region.
Therefore the proper abundances for the heavy elements are those
derived from the forbidden lines adopting the $t^2$ value derived from
the combination of $T_e$(Balmer) and $T_e$([O~III]), for those PNe
with most of their oxygen in the O$^{++}$ stage. If these abundances
are in agreement with those derived from recombination lines it means
that the nebula is chemically homogeneous.

In Table 3 we present temperature values for six very bright PNe
derived from four different sets of He~I recombination lines and
compare the results derived by Zhang et al. (2005) with those derived
by Peimbert, Luridiana, \& Torres-Peimbert (1995b). The best
comparison between $T_e$(Balmer) and $T_e$(He~I) is provided by
objects where most of the He is in the form of He$^+$, because then
the He~I lines and the H~I lines originate in the same volume. For
NGC~6572, NGC~6803, and NGC~7009 most of the He is in the He$^+$
stage, and the $T_e$(Balmer) and $T_e$(He~I) values are practically
the same when $T_e$(He~I) is derived from $\lambda\lambda$ 3889, 4471,
and 7065, contrary to the results derived from $\lambda$ 6678 and
$\lambda$ 7281, and in favor of a homogeneous chemical composition for
these three objects. Alternatively the results derived from
$\lambda\lambda$ 3889, 4471, and 10830 are intermediate between those
derived from $\lambda\lambda$ 3889, 4471 and 7065, and those derived
from $\lambda$ 6678 and $\lambda$ 7281 providing support for the
two-abundance nebular model. For the other three PNe the three sets of
He~I lines clearly indicate that $T_e$(He~I) is smaller than
$T_e$(Balmer), but in these three PNe a large fraction of He is in the
He$^{++}$ region where a higher temperature is expected and where a
fraction of the Balmer line emission originates, particularly in the
case of Hu 1-2.  The differences in the $T_e$(He~I) values derived
from different sets of He~I recombination lines need to be sorted
out.

\begin{table}\def~{\hphantom{0}}
  \begin{center}
  \caption{Electron Temperatures (K)}
  \label{tab:ne}
  \begin{tabular}{lccccc}\hline
Object   & $T_e$([O~III]) & $T_e$(Balmer)  & $T_e$(He~I)$^a$ & $T_e$(He~I)$^b$& $T_e$(He~I)$^c$\\
    \hline
NGC~6572 & $10 500\pm300$ & $10300\pm1000$ & $ 9800\pm600  $ & $7100\pm500$  & $ 8690\pm1200$ \\
NGC~6803 & $10 000\pm300$ & $ 8500\pm 400$ & $ 8500\pm500  $ & $6900\pm400$  & $ 5000\pm1100$ \\
NGC~7009 & $10 000\pm300$ & $ 7200\pm 400$ & $ 8000\pm400  $ & $6800\pm400$  & $ 5040\pm 800$ \\
NGC~7027 & $13 000\pm300$ & $12000\pm 400$ & $10000\pm600  $ & $8200\pm600$  & $10360\pm1100$ \\
NGC~7662 & $13 000\pm300$ & $12200\pm 600$ & $ 9500\pm600  $ & $9200\pm700$  & $ 7690\pm1650$ \\
Hu~1-2   & $18 900\pm300$ & $20000\pm1200$ & $12900\pm600^d$ & ...           & $11500\pm1500$ \\
  \hline
  \end{tabular}
  
$T_e$([O~III]), $T_e$(He~I)$^{a,b,d}$ Peimbert et al.(1995b); $T_e$(Balmer) Zhang et al.
(2004); $T_e$(He~I)$^c$ Zhang et al.(2005); a) 3889,4471,7065; b) 3889,4471,10830;
c) 6678,7281; d) 4471,5876,6678.

  \end{center}
\end{table}

\section{Conclusions}

We consider that our knowledge on the density and temperature distributions and
on the chemical composition of PNe will increase considerably from the study of
the four following problems.

The $N_e$(O~II) values derived from the O~II lines of multiplet 1, like those
presented in Table 3, need to be determined again based on the atomic physics
computations by Bastin \& Storey (2006). Objects with small He$^0$ and
He$^{++}$ fractions and with most of their O in the O$^{++}$ ionization stage
will be particularly useful, those objects of this group with $N_e$(O~II) 
$\sim N_e$(Balmer) will be chemically homogeneous, while those with $N_e$(O~II) $>$ 
$N_e$(Balmer) will be chemically inhomogeneous.

The idea that there are high density He$^+$ regions embedded in low density H
rich material might be tested by deriving $T_e$(He~I), $N_e$(He~I), and
$\tau$(3889) based on accurate measurements of at least 10 different He~I
lines in relatively low density PNe without substantial
He$^0$ and He$^{++}$ regions. Those objects with $N_e$(He~I) $\sim N_e$(Balmer) 
will be chemically homogeneous, while those with $N_e$(He~I) $> N_e$(Balmer) 
will be chemically inhomogeneous.

There are at least seven possible mechanisms as sources of temperature
variations. From the theoretical side, it is important to model the
signature of each process on the temperature and density
distributions. While, from the observational side, the combination of
3D kinematical models with high spectral resolution data, like those
presented by Barlow et al. (2006), might permit to derive
temperature and density distributions and consequently to single out 
the main mechanism responsible for the temperature variations in a given object.

Finally accurate H, He, C, and O abundances of H-rich central stars, like those
obtained by Georgiev et al. (2006) for NGC~6543, are needed to compare them with
those nebular abundances derived from permitted and forbidden lines.

We are grateful to the SOC for an outstanding meeting and to Roberto
M\'endez and the LOC for their warm hospitality. We would also like to
acknowledge partial support received from CONACyT grant 46904.


\begin{thebibliography}{}
     
\bibitem[]{}{Asplund, M., Grevesse, N. \& Sauval, A. J.} 2005, in: ASP Conf. Ser. 336, \textit{Cosmic
 Abundances as Records of Stellar Evolution and Nucleosynthesis}, ed. F. N. Bash
 \& T.G. Barnes(San Francisco:ASP)

\bibitem[]{}{Barlow, M. J., Hales, A. S., Storey, P. J., Liu, X.-W., Tsamis, Y. G. \&
 Aderin, M. E.} 2006, these proceedings
 
\bibitem[]{}{Bastin, R. J. \& Storey, P. J.} 2006, these proceedings


\bibitem[]{}{Carigi, L.} 2003, {\it MNRAS}, 339, 825

\bibitem[]{}{Carigi, L., Peimbert, M., Esteban, C. \& Garc\'{\i}a-Rojas, J.} 2005, {\it ApJ}, 623, 213

\bibitem[]{}{Esteban, C.} 2002, {\it RevMexAASC}, 12, 56
 
\bibitem[]{}{Esteban, C., Garc\'{\i}a-Rojas, J., Peimbert, M., Peimbert, A., Ruiz, M. T.,
 Rodr\'{\i}guez, M. \& Carigi, L.} 2005, {\it ApJ} 618, L95
 
\bibitem[]{}{Georgiev, L., Hillier, D. J., Richer, M. \& Arrieta, A.} 2006, these proceedings

\bibitem[]{}{Guerrero, M. A., Chu, Y.-H. \& Gruendl, R. A.} 2006, these proceedings

\bibitem[]{}{Guerrero, M. A., Chu, Y.-H., Gruendl, R. A. \& Meixner, M.}  2005, {\it A\&A} 430, L69

\bibitem[]{}{Gorny, S. K. \& Tylenda, R.} 2000, {\it A\&A} 362, 1008

\bibitem[]{}{Hazard, C., Terlevich, R., Morton, D. C., Sargent, W. L. W. \& Ferland, G.} 1980, {\it Nature}
285, 463

\bibitem[]{}{Huggins, P. J. \& Frank, A.} 2006, these proceedings

\bibitem[]{}{Jacoby, G. H.} 1979, {\it PASP} 91, 754

\bibitem[]{}{Jacoby, G. H. \& Ford, H. C.} 1983, {\it ApJ}, 266, 298

\bibitem[]{}{Kleinman, S. J., Harris, H. C. , Eisenstein, D. J., Liebert, J. et al.} 2004, {\it ApJ}, 607, 426
 
\bibitem[]{}{Liu, X.-W.} 2003, in: \textit{Planetary Nebulae: Their Evolution and Role in the Universe}
 S. Kwok, M. Dopita \& R. Sutherland (eds.), (San Francisco:ASP), p. 33>9
 
\bibitem[]{}{Liu, X.-W.} 2006, these proceedings

\bibitem[]{}{Liu, X.-W. \& Danziger, I. J.} 1993, {\it MNRAS} 263, 256

\bibitem[]{}{Liu, X.-W., Luo, S.-G., Barlow, M. J., Danziger, I. J. \& Storey, P. J.} 2001,
{\it MNRAS} 327, 141

\bibitem[]{}{Luo, S. G. \& Liu, X.-W.} 2003, in: \textit{Planetary Nebulae: Their Evolution and Role in the Universe}
 S. Kwok, M. Dopita \& R. Sutherland (eds.), (San Francisco:ASP), p. 393

\bibitem[]{}{Manchado, A., Pottasch, S. R. \& Mampaso, A.} 1988, {\it A\&A} 191, 128

\bibitem[]{}{Marigo, P., Bressan, A. \& Chiosi, C.} 1996, {\it A\&A} 313, 545

\bibitem[]{}{Mathis, J. S.} 1976, {\it ApJ} 207, 442

\bibitem[]{}{Matteucci, F.} 2006, astro-ph/0603820

\bibitem[]{}{Medina, S., Pe\~na, M. \& Stasi\'nska, G.} 2006, {\it RevMexAA} 42, 53

\bibitem[]{}{Peimbert, A. \& Peimbert, M.} 2005,  {\it RevMexAAS} 23, 9

\bibitem[]{}{Peimbert, M}. 1967, {\it ApJ} 150, 825

\bibitem[]{}{Peimbert, M.} 1971, \textit{Bol. Obs. Tonantzintla y Tacubaya}, 6, 29

\bibitem[]{}{Peimbert, M. \& Costero, R.} 1969, \textit{Bol. Obs. Tonantzintla y Tacubaya}, 5, 3

\bibitem[]{}{Peimbert, M., Luridiana, V. \& Torres-Peimbert, S.} 1995b, {\it RevMexAA} 31, 147

\bibitem[]{}{Peimbert, M., Peimbert, A., Ruiz, M. T. \& Esteban, C.} 2004, {\it ApJS} 150, 431

\bibitem[]{}{Peimbert, M., Storey, P. J. \& Torres-Peimbert, S.} 1993, {\it ApJ} 414, 626

\bibitem[]{}{Peimbert, M., Torres-Peimbert, S. \& Luridiana, V.} 1995a, {\it RevMexAA} 31, 131

\bibitem[]{}{Rola, C. \& Stasi\'nska, G.} 1994, {\it A\&A} 282, 199

\bibitem[]{}{Rowlands, N., Houck, J. R. \& Herter, T.} 1994, {\it ApJ} 427, 867

\bibitem[]{}{Ruiz, M. T., Peimbert, A., Peimbert, M. \& Esteban, C.} 2003, {\it ApJ} 595, 247

\bibitem[]{}{Sandin, C., Sch\"onberner, D., Roth, M. M., Steffen, M., Monreal-Ibero, A. B\"ohm, P.
\& Tripphahn, U.} 2006, these proceedings

\bibitem[]{}{Stasi\'nska, G. \& Szczerba, R.} 2001, {\it A\&A} 379, 1024

\bibitem[]{}{Storey, P. J.} 1994, {\it A\&A} 282, 999

\bibitem[]{}{Torres-Peimbert, S. \& Peimbert, M.} 2003, in: \textit{Planetary Nebulae: Their Evolution and Role
 in the Universe} S. Kwok, M. Dopita \& R. Sutherland (eds.), (San
 Francisco:ASP), p. 339

\bibitem[]{}{Tylenda, R.} 2003, in: \textit{Planetary Nebulae: Their Evolution and Role
 in the Universe} S. Kwok, M. Dopita \& R. Sutherland (eds.), (San
 Francisco:ASP), p. 451

\bibitem[]{}{Wesson, R. \& Liu, X.-W.} 2004, {\it MNRAS} 351, 1026

\bibitem[]{}{Wesson, R., Liu, X.-W. \& Barlow, M. J.} 2003, {\it MNRAS} 340, 253

\bibitem[]{}{Zhang, Y., Liu, X.-W., Wesson, R., Storey, P. J., Liu, Y. \& Danziger, I. J.} 2004, {\it MNRAS} 351, 935

\bibitem[]{}{Zhang, Y., Liu, X.-W.,Liu, Y. \& Rubin, R. H.} 2005, {\it MNRAS} 358, 457

\end{thebibliography}
\end{document}